\documentstyle[aps,prl,epsfig,twocolumn]{revtex}

\renewcommand{\v}[1]{{\mathbf #1}}        % \mathbf f"ur Vektoren
\renewcommand{\rho}{\varrho}              % sch"oneres \rho
\renewcommand{\phi}{\varphi}              % sch"oneres \phi
\newcommand{\vf}{v_{\mathrm F}}             % Fermi velocitiy
\newcommand{\Ef}{E_{\mathrm F}}             % Fermi velocitiy
\newcommand{\omcy}{\omega_{\mathrm c}}      % cyclotron frequency
\newcommand{\omer}{\omega_{\mathrm tot}}      % cyclotron frequency
\newcommand{\intpp}{\int_{-\pi}^{\pi}\frac{d\phi'}{2\pi}\,} % Integral über phi'
\newcommand{\intp}{\int_{-\pi}^{\pi}\frac{d\phi}{2\pi}\,} % Integral über phi
\newcommand{\tautr}{\tau_{\mathrm tr}}       % transport scattering time
\newcommand{\cf}{{\mathrm{CF}}}         % Index Electron

\draft % Wenn man das wegnimmt, verschwinden die pacs-Nummern... 

\begin{document}
%\twocolumn[{
\title{
Positive Magnetoresistance of Composite Fermions in Laterally Modulated 
Structures 
}
\author{ 
S. D. M. Zwerschke and R. R. Gerhardts}
\address{
Max-Planck-Institut f\"ur Festk\"orperforschung, Heisenbergstra\ss e 1, 
D-70569 Stuttgart, Germany \\}

\date{\today}
\twocolumn[
\setlength{\columnwidth}{18cm}%
\csname@twocolumnfalse\endcsname %taeuscht \maketitle vor, wir seien im onecolum

\maketitle

\begin{abstract}
  Adopting the mean-field composite fermion picture, we describe the
  magneto-transport properties of a two-dimensional electron gas with
  laterally modulated density around filling factor $\nu=1/2$. 
  The occurrence of a strong positive magnetoresistance at low effective 
  magnetic fields as well as Weiss oscillations, which were observed in recent
  experiments in such systems,  can be explained within a semi-classical
  Boltzmann equation approach, provided one goes beyond a second order
  approximation in the modulation strength.

\end{abstract}
]

%\maketitle

%\begin{abstract}
%  Adopting the mean-field composite fermion picture, we describe the
%  magneto-transport properties of a two-dimensional electron gas with
%  laterally modulated density around filling factor $\nu=1/2$. 
%  The occurrence of a strong positive magnetoresistance at low effective 
%  magnetic fields as well as Weiss oscillations, which were observed in recent
%  experiments in such systems,  can be explained within a semi-classical
%  Boltzmann equation approach, provided one goes beyond a second order
%  approximation in the modulation strength.

%\end{abstract}
%\pacs{PACS numbers: 71.10.Pm,73.40.Hm,73.20.Dx}

Transport properties of strongly correlated two-dimensional electron systems
(2DES) in high magnetic
fields, where the lowest Landau level is about half filled,
 seem to be surprisingly
well described by the ``Composite Fermion'' (CF) picture. 
This model is motivated by a singular Chern-Simons gauge-field transformation 
that maps the electron system near filling factor $\nu=1/2$
onto a metal of quasi-particles (the CFs). They can be interpreted as
electrons to which two magnetic flux quanta have been 
attached\cite{Jain89:199}. %,Halperin97:225}  
In the usual mean-field approximation, the CFs are 
considered as having the same density $n_{\mathrm{CF}}=n_{\mathrm{el}}\equiv
n$ as the electrons and as moving in the weak effective magnetic field 
$B_{\mathrm{eff}}=B-2\Phi_0 n$, 
where $B$ is the external applied magnetic flux density and $\Phi_0=h/e$ the
flux quantum. Halperin, Lee, and Read\cite{Halperin93:7312}  predicted
 that, at low temperatures, these CFs fill in momentum space the
 spin-polarized  states 
within a well defined Fermi circle of radius $k_{\mathrm F}=\sqrt{4\pi n}$.
Therefore, one expects that transport properties, which are governed by
elastic scattering at the Fermi energy, can be 
calculated without considering explicitly the mutual interaction between 
the CFs, by analogy with the Landau theory of the electron Fermi liquid in zero
and weak magnetic fields.
Indeed, interesting features of the magneto-resistance of 2DESs (near
$\nu=1/2$) in lateral superlattices, such as two-dimensional anti-dot lattices
or  one-dimensional (1D)  superlattices, created
either dynamically by the application of surface acoustic waves or statically
by surface etching, have been reproduced
astonishingly well by calculations describing the CFs as
classical, non-interacting particles moving in suitable effective
fields\cite{Fleischmann96:167,Mirlin97:3717,vonOppen98:4494}. 

Here, we focus on static 1D density modulations and  compare
experimental dc-transport measurements \cite{Smet98:4538} with calculations 
based on the linearized Boltzmann equation (LBE). The purpose of the present
letter is to  
demonstrate that {\em all} characteristic features of the magnetoresistance 
curve near $B=B_{1/2} \equiv 2\Phi_0 n$ can be
 reproduced in the quasi-classical CF picture, provided one 
incorporates {\em all modulating fields} and solves the LBE 
{\em beyond} the Beenakker-type approximation (BA)\cite{Beenakker89:2020},
which allows an analytical solution and has been employed in previous
work\cite{Mirlin97:3717,vonOppen98:4494,Mirlin98:1070}. 
These features are: (i) a pronounced 
V-shaped minimum near $B_{1/2}$, explained by ``channeled orbits'' which are
omitted in the BA, (ii) shoulders
or minima related to commensurability effects (Weiss oscillations due to
drifting cyclotron orbits), (iii) a steep increase of the resistance with
$|\bar{B}_{\mathrm{eff}}|$ between the structures caused by commensurability effects
and the oscillatory structures due to the fractional quantum Hall effect, and
(iv) an asymmetry in the slope and magnitude of these steep resistance flanks
with respect to $\bar{B}_{\mathrm{eff}}=0$, due to 
interference effects between the direct electrostatic density 
modulation and the induced modulation of $B_{\mathrm{eff}}$. The material
parameters needed to achieve agreement with the experiment around $\nu=1/2$
are consistent with
those needed to explain the commensurability effects at low magnetic fields
($B < 0.5\,$T). 

Recent work by Mirlin \emph{et al.}\cite{Mirlin98:1070}, based on an analytical
solution of the LBE within the BA for a special model of anisotropic
scattering\cite{Mirlin97:3717}, obtained reasonable results for the features
(ii) and (iii), but not for (i) and (iv). Their approximation
misses relevant  physics  
at low $\bar{B}_{\mathrm{eff}}$, where the CF picture is expected 
to apply best, and gives a rather poor fit to the experiment\cite{Smet98:4538}
for very small $|\bar{B}_{\mathrm{eff}}|$ (see Fig~\ref{mod_400} below).

To describe the resistance for $B < 0.5\,$T, we follow
Ref.~\cite{Menne98:1707} for pure electric modulation, and treat the 2DES as a
degenerate Fermi gas of non-interacting particles, with charge $-e$ and
effective mass $m^*_{\mathrm el}$, average density $\bar{n}$,
Fermi energy $\Ef^{\mathrm el} =\bar{n}/D_0^{\mathrm el}$, and
density of states $D_0^{\mathrm el}=m^*_{\mathrm el}/(\pi\hbar^2)$. The
etching of grooves into the surface of GaAs-heterostructures is
assumed to produce an external electrostatic
potential energy $V^{\mathrm ext}(x)= V_0^{\mathrm ext} \cos qx$ of period
$a=2\pi /q$ in the plane of the 2DES. It is screened 
by the 2DES and this leads within the Thomas-Fermi approximation to a potential
energy $V^{\mathrm{el}}(x)=V_0^{\mathrm{el}}  \cos qx$ 
with $V_0^{\mathrm el} = V_0^{\mathrm{ext}} / [1+ 2/(q a_B^*)]$, where  
$a_B^* = 1/[ \pi e^2 D_0^{\mathrm el}/\kappa]$ is the 
effective Bohr radius. For GaAs, $\kappa=12.4$ and $a_B^* \approx 10\,$nm $\ll
a \sim 400\,$nm. Consequently, the relative modulation strength is
$\epsilon_{\mathrm{el}} 
= V_0^{\mathrm el} / \Ef^{\mathrm el} \approx V_0^{\mathrm ext} \kappa a/(e^2 
\bar{n})$, and the modulated electron density is 
$n(x)=\bar{n}[1- \epsilon_{\mathrm el}\cos qx]$, independent of $m^*_{\mathrm{el}}$.

To describe the resistance of CFs, we modify the approach of Ref.~\cite{Menne98:1707}.
We treat the CF system as a degenerate
Fermi gas of non-interacting particles, with charge $-e$, effective mass
$m^* _{\mathrm CF}$ \cite{Fleischmann96:167}, average density $\bar{n}$, Fermi energy 
$\Ef^{\mathrm CF} = \bar{n}/D_0^{\mathrm CF}$, and density of states 
$D_0^{\mathrm CF} = m^*_{\mathrm CF} /(2 \pi\hbar ^2)$, which obeys Newton's
equation,   
$m^* _{\mathrm CF}\dot{\v{v}}=-e[\v{F}_{\mathrm{eff}}+\v{v}\times
(B_{\mathrm{eff}}\v{e}_z)]$, with effective electric and magnetic fields.
We use the LBE to calculate the response to an
external homogeneous electric field $\v{E}^{(0)}$.
In the absence of $\v{E}^{(0)}$, $\v{F}_{\mathrm eff}=\nabla V(x)/e$ is
determined by the screened modulation potential, which we parametrize as 
$V(x)=\epsilon _{\mathrm CF} \Ef ^{\mathrm CF} \cos(qx)$.
Since the static Thomas-Fermi screening of the CFs should be equivalent to
that of electrons (with  $ D_0^{\mathrm  CF}>D_0^{\mathrm{el}}$ instead of $ D_0^{\mathrm
  el}$), we expect $\epsilon _{\mathrm CF} \approx \epsilon_{\mathrm el}$, 
so that
the modulated
density  $n(x) = \bar{n}[1- \epsilon_{\mathrm{CF}} \cos qx]$
for $B$ near $B_{1/2}$ is the same as that for small $B$ within this approximation.
This yields the effective magnetic field
$B_{\mathrm{eff}}(x)=B-2\Phi_0n(x)= \bar{B}_{\mathrm{eff}}
+\delta B_{\mathrm{eff}}(x)$, where
$\delta B_{\mathrm{eff}}(x)= \epsilon_{\mathrm CF} \, B_{1/2}
\cos qx$ is the magnetic modulation.

The linear response to $\v{E}^{(0)}$ 
is carried by CFs at the Fermi edge, 
$(m^*_{\mathrm CF}/2){\v{v}}^2+V(x)=\Ef^{\mathrm CF}$. 
Due to
the translational invariance in $y$ direction, the (suitably scaled)
zero-temperature distribution function  $\Phi(x,\phi)$ \cite{Menne98:1707}
depends only on two variables, the position
$x$ and the polar angle $\phi$ of the velocity
$\v{v}(x,\phi)=v(x)(\cos\phi,\sin\phi)$, with $v(x)=\vf[1-V(x)/\Ef^{\mathrm
  CF} ]^{1/2}$, $\vf= \hbar k_{\mathrm{F}}^{\mathrm CF} /  m^*_{\mathrm
  CF} $, and $ k_{\mathrm{F}}^{\mathrm CF} = [ 4 \pi\, \bar{n}]^{1/2}$. We
normalize $\Phi(x,\phi)$ so that the  current density ${\v{j}}(x)$ reduces to  
\begin{equation}
  \label{current_density}
  {\v{j}}(x)=e^2D_0^{\mathrm CF} \intp \v{v}(x,\phi)\Phi(x,\phi).
\end{equation}
A current of CFs implies motion of flux tubes and produces the Chern-Simon
electric field \cite{Simon98:91} 
\begin{equation}
  \label{Chern-Simons-Electric}
  {\v{E}}^{\mathrm{CS}}=-(2h/e^2)\,\v{j}\times\v{e}_z,
\end{equation}
which, in the linearized Boltzmann equation,
\begin{equation}
  \label{Boltzmann_modulated}
  {\mathcal D}\Phi-{\mathcal C}[\Phi]=\v{v}(x,\phi)
  \left(\v{E}^{(0)}+\v{E}^{\mathrm{CS}}(x)\right),
\end{equation}
adds to the external driving field $\v{E}^{(0)}$. Therefore a self-consistent
solution of Eqs.~(\ref{current_density})-(\ref{Boltzmann_modulated}) is necessary.
The drift operator 
\begin{equation}
  \label{Drift_Term_1}
  {\mathcal D}=v(x)\cos\phi\,\partial_x + \big(\omcy + 
  \omega_{\mathrm  m}(x)+\omega_{\mathrm e}(x,\phi)\big)\partial_\phi,
\end{equation}
contains the average value and the periodic component of the effective
magnetic field via $\omcy=e\bar{B}_{\mathrm{eff}}/m^*_{\mathrm{CF}}$, and 
$\omega_{\mathrm m}(x)=e\delta B_{\mathrm{eff}}(x)/m^*_{\mathrm{CF}}$,
respectively.
The electric modulation enters in 
$\omega_{\mathrm{e}}(x,\phi)=-\sin\phi\,dv/dx$ and $v(x)$
\cite{Menne98:1707}. The collision operator is taken to be the 
same as for the unmodulated system and is written in the form
\begin{equation}
  \label{scattering_operator}
  {\mathcal C}[\Phi]=\frac{1}{\tau}\intpp\,P(\phi'-\phi)
  \left[\Phi(x,\phi')-\Phi(x,\phi)\right],
\end{equation}
where the differential cross section is parametrized by a relaxation time
$\tau$ and a dimensionless kernel $P(\phi)$. For actual calculations we 
take $ P(\phi)=b+(1-b)P_p(\phi)$ with $0\le b\le 1$ and
\begin{equation}
  \label{in_kernel}
  P_p(\phi)=[(2^pp!)^2/(2p)!]\cos^{2p}(\phi/2) \, ,
\end{equation}
 which has the finite Fourier expansion  
$ P(\phi)=\sum_{n=0}^p\gamma_n\,\cos n\phi$ with $\gamma_0=1$, and 
$\gamma_n=2(1-b)(p!)^2/((p+n)!(p-n)!)$ for $n\ge1$. 
For $b=1$ and $p=0$, $P(\phi)$ describes isotropic scattering. For $b<1$ the
fraction $(1-b)$ of the total scattering cross section is due to anisotropic
scattering. With increasing $p$, $P_p(\phi)$
is increasingly stronger peaked in forward direction.

To solve Eqs.~(\ref{current_density})-(\ref{Boltzmann_modulated}), we 
calculate the distribution function
of the homogeneous, unmodulated CF system first. This yields the resistivity 
tensor $\hat{\rho}^{\mathrm{h}}$ with components 
given by  $\rho_{xx}^{\mathrm{h}}=
\rho_{yy}^{\mathrm{h}}=\rho_0^{\mathrm{CF}}$ and
$\rho_{xy}^{\mathrm{h}}=-\rho_{yx}^{\mathrm{h}}=
\omer\tautr\rho_0^{\mathrm{CF}}$, where  
$\rho_0^{\mathrm{CF}}=m^*_{\mathrm{CF}}/(e^2\bar{n}\tautr)$ and
$\tautr=\tau/(1-\gamma_1/2)$ is the relevant CF transport scattering time.
 Although the CFs move in the effective magnetic field $B-B_{1/2}$,
 the Hall resistance $\rho_{xy}^{\mathrm{h}}$ is determined by the 
 cyclotron frequency  $\omer=\omcy+\omega_{1/2}$ for the total magnetic field
 $B$,   with $\omega_{1/2}=eB_{1/2}/m^*_{\mathrm{CF}}$, due to the inclusion
 of the Chern-Simons electric field. Introducing the mean free path
 $\lambda_{\mathrm{CF}} =\tautr \, \hbar
 k_{\mathrm{F}}^{\mathrm{CF}}/m^{*}_{\mathrm{CF}} $, we obtain  
$\rho_0^{\mathrm{CF}}=\hbar k_{\mathrm{F}}^{\mathrm{CF}}
/(e^2\bar{n} \lambda_{\mathrm{CF}})$  and 
$\rho_{xy}^{\mathrm{h}}=B/(\bar{n}e)$, so that
the resistivity tensor is independent of $m^*_{\mathrm{CF}}$, and the Hall
resistance is the same as that of the 2DES.
Comparing with the corresponding Drude formulas for the 2DES at low $B$, we
see that  the ratio of the longitudinal resistances,
$\rho_0^{\mathrm{CF}}/\rho_0^{\mathrm{el}}= \sqrt{2}\lambda_{\mathrm{el}}/
\lambda_{\mathrm{CF}}$, directly reflects the ratio of the corresponding mean
free paths. 

The macroscopic resistivity tensor
$\hat{\rho}$ of the modulated system, relates the spatial average 
$\langle\v{j}(x)\rangle$ of the current density $\v{j}(x)$ to the driving
field, $\hat{\rho} \, \langle\v{j}(x)\rangle=\v{E}^{(0)}$.
Exploiting the solution for the homogeneous system and
the continuity equation, which implies that $j_x(x)$ for the modulated
system is independent of $x$, it can be shown that 
$\hat{\rho}$ differs from $\hat{\rho}^{\mathrm{h}}$ only in its $xx$-component.
 With some formal but exact
manipulations (similar to those of ref.~\cite{Menne98:1707}) one finds
\begin{equation}
  \label{resistivity}
  (\rho_{xx}-\rho_0^{\mathrm{CF}})/\rho_0^{\mathrm{CF}}=G/[1-G/(1+(\omer\tautr)^2)].
\end{equation}
In this equation
\begin{equation}
  \label{G}
  G=\frac{2 \tautr}{v_{\mathrm{F}}^2 }\left\langle\intp 
  g(x,\phi)\chi(x,\phi)\right\rangle,
\end{equation}
where $g(x,\phi)=\vf^2\,(dV/dx)/(2\Ef^{\mathrm{CF}}) +\omega_{\mathrm{m}}v_y$
is of first order 
in the modulation strength $\epsilon_{\mathrm{CF}}$, and
$\chi(x,\phi)$ is the solution of the modified Boltzmann equation
\begin{equation}
  \label{dgl}
  {\mathcal D}\chi-{\mathcal C}[\chi]=g(x,\phi)+
  \omega_{1/2}v_y\left(j_x/j_x^{\mathrm{h}}-1\right).
\end{equation}
The basic differences between these results for the CF system and those for an
electron system \cite{Menne98:1707} derive from inclusion of the Chern-Simons
electric field in Eq.~(\ref{Boltzmann_modulated}):
It is responsible for the occurence of $\omer\propto B$ in the denominator of
Eq.~(\ref{resistivity}), instead of $\omcy\propto B-B_{1/2}$, and 
the last term in the right hand side of Eq.~(\ref{dgl}). 
This last term can be expressed
in terms of $G$, $j_x/j_x^{\mathrm{h}}-1=-G/(1+(\omer\tautr)^2)$, and requires
a self-consistent solution of Eqs.~(\ref{resistivity})-(\ref{dgl}). 
Inserting $\tautr = \lambda_{\mathrm{CF}} / \vf$ into Eq.~(\ref{G}) makes $G$
and $\rho_{xx}$ independent of $m^*_{\mathrm{CF}}$, as it should be.

Numerical solution of Eqs.~(\ref{resistivity})-(\ref{dgl}) for the CF system
and of the corresponding equations of Ref.~\cite{Menne98:1707} for the
electron system allows to calculate the resistance of the modulated sample
for strong applied fields $B\approx B_{1/2}$ and in the low $B$ regime,
respectively.
To compare with experiment, we need to choose a reasonable set
of parameters. Whereas in both $B$ regimes
the average density $\bar{n}$ and the modulation period
$a$ are the same, and also the modulation strengths $\epsilon_{\mathrm{CF}}
\approx \epsilon_{\mathrm{el}}$ should be similar, the parameters
  characterizing the collision operator (in our model $\tau$, $b$ and $p$) are
  expected to be different. To reproduce the characteristics of the
  Weiss oscillations observed at small $B$ \cite{Smet98:4538},  we have to
  assume a very anisotropic differential cross section with
a sharp peak in forward direction and a zero-$B$ resistance of only a
few $\Omega$ ($\lambda _{\mathrm{el}} \sim 50\, \mu$m). This is
reasonable since these high-mobility samples have a spacer thickness of about 
$60\,$nm, so that  the donors produce a smooth random potential in the plane
of the 2DES that scatters the electrons predominantly under small angles.
The CFs, on the other hand, are scattered not only by these small-amplitude
donor-induced random 
potential fluctuation, but also -- and predominantly -- by the large-amplitude
random fluctuations of the 
effective magnetic field that originates from the donor-induced density
fluctuation. This results in $\lambda_{\mathrm{CF}} \ll \lambda_{\mathrm{el}}$
and a less pronounced forward scattering for the CFs. 

Figure \ref{mod_400} shows resistivity data for a sample with 
density $\bar{n}=1.82\cdot10^{11}\ \mathrm{cm}^{-2}$ and period
$a=400$ nm near filling factor $\nu=1/2$ \cite{footnote}. The thick short-dashed line is the
experiment of Ref.~\cite{Smet98:4538}. The thin dash-dotted line is the
theoretical curve using the approximations of Ref.~\cite{Mirlin98:1070}, 
based on the Beenakker-type approximation
(BA) mentioned above. 
It neglects the modulation effects in the drift operator
$\mathcal{D}$ of the LBE, that causes a resistance correction
quadratic in $\epsilon_{\mathrm{CF}}$.
Furthermore, the direct effect of the electric modulation on the CFs was
ignored in this fit. Neglecting this  direct effect too and using the BA,
we can closely reproduce this 
symmetric curve of Ref.~\cite{Mirlin98:1070} using
the same values for $\bar{n}$, $\rho_0^{\mathrm{CF}}$,
$\epsilon_{\mathrm{CF}}$, and using $b=0$, $p=6$ for our model of
the scattering cross section, 
Eqs.~(\ref{scattering_operator}) and (\ref{in_kernel}). If we include the
direct electric modulation in addition to the magnetic one, we obtain the
asymmetric 
thin dotted line. For small $|\bar{B}_{\mathrm{eff}}|$, this again closely
reproduces the results of Ref.~\cite{Mirlin98:1070}, but at large
$|\bar{B}_{\mathrm{eff}}|$ the curve shows a distinct asymmetry, similar to the
experimental curve, and similar to the asymmetry obtained for mixed electric
and magnetic modulations in 2DESs at low $B$
\cite{Ye95:3013,Gerhardts96:11064}. Solving
Eqs.~(\ref{resistivity})-(\ref{dgl}) for the same set of parameters but
without any approximation of the drift 
operator, we obtain the thin solid curve of Fig.~\ref{mod_400}. This shows
that approximating the drift operator $\mathcal{D}$ by the one of the
homogeneous system is insufficient 
for $|\bar{B}_{\mathrm{eff}}|<0.5\,$T. Physically the BA means that the forces
due to the modulation fields are neglected against the Lorentz force due to
the average magnetic field $\bar{B}_{\mathrm{eff}}$. This omits the effect of
channeled orbits, i.e. of ``snake orbits'' \cite{Mueller92:385} along the
lines of vanishing $B_{\mathrm{eff}}(x)$ which, similar to the ``open orbits''
\cite{Beton90:9229} in the case of a pure electric modulation, cause a
pronounced positive low-field magneto-resistance
\cite{Beton90:9229,Menne98:1707}. 
Since $B_{1/2}=15\,$T \cite{Smet98:4538}, the effective magnetic modulation
has an amplitude $\epsilon_{\mathrm{CF}} B_{1/2} \approx 0.45\,$T, so that the
effect of channeled trajectories should be important for
$|\bar{B}_{\mathrm{eff}}|<0.45\,$T, and a second order approximation in
$\epsilon_{\mathrm{CF}}$ is not adequate for the $\bar{B}_{\mathrm{eff}}$ values
shown in Fig.~\ref{mod_400}.

Comparing the thin solid curve with the experimental one
suggests that the pronounced V-shaped resistance minimum at
$\bar{B}_{\mathrm{eff}}=0$ seen in the experiment is indeed due to CFs on channeled
orbits. 
To achieve a better quantitative agreement, we choose more realistic
parameters. First we take  $b=0.1$ and $p=2$, since then our model
(\ref{in_kernel}) 
approximates closely (apart from the $\phi =0$ divergence) the 
differential cross section  derived by Aronov \emph{et
  al.}\cite{Aronov94:16609} for random magnetic field scattering. This
choice  allows a good fit of the
resistance near $\nu=1/2$ for all samples available to us, indicating the same
scattering mechanism in those samples. Then, assuming a density modulation of
about 3\%, we have to take $\rho_0^{\mathrm{CF}}=650\ \Omega$
($\lambda_{\mathrm{CF}}\approx 1.3\,a$), i.e.\ a larger value than the
resistance of the unmodulated reference sample ($\rho_0^{\mathrm{CF}}=270\
\Omega$) taken in Ref.~\cite{Mirlin98:1070}. This yields the thick
solid curve from  the exact solution of the LBE, which  reproduces nicely all
features of the experimental data.
 The larger $\rho_0^{\mathrm{CF}}$ value  is reasonable since the
etching procedure, in addition to the intended modulation, introduces
unintended defects which increase the resistance.

%The insert of Fig.~\ref{mod_400} shows the measured (dashed curve) low-field
%resistance for this sample, together with a calculation based on
%Ref.~\cite{Menne98:1707}.
%We need extremely pronounced forward scattering
%($b=0$ and $p=15$) in order to reproduce the small measured resistance
%($\rho_0^{\mathrm{el}}=4.5\ \Omega$, i.e.\ $\lambda_{\mathrm{el}}\approx
%120a$) and simultaneously the strong suppression of Weiss
%oscillations at $B<0.05\,$T. The good agreement is of course limited by the
%Shubnikov-de Haas oscillations which in these 
%high-mobility samples (at $T=50\,$mK) set in at $B>0.2\,$T.

For the sample discussed so far ($a=400$ nm), the mean free path of the CFs is
only marginally larger than the modulation period, $\lambda_\cf\approx 1.3 a$,
so that commensurability effects are  visible only as weak
shoulders (thick lines of Fig.~\ref{mod_400}). To see these effects more
clearly, we have investigated a sample with a smaller modulation period,
$a=275$ nm, so that we expect $\lambda_\cf \approx 1.9 a$. Experimental \cite{SmetPrivat} and
theoretical results (with the same anisotropy parameters $b$ and $p$ as in
Fig.~\ref{mod_400}) are plotted in Fig.~\ref{smet_285}. Indeed, the
commensurability effects near $|\bar{B}_{\mathrm{eff}}|=0.5\,$T
are more pronounced. Again, the low field resistance (inset of
Fig.~\ref{smet_285}) and that near 
$\nu=1/2$ can be reproduced theoretically with a consistent
set of parameters. 

In conclusion, our numerical solution of the linearized Boltzmann equation for composite
Fermions without the
Beenakker-type approximation used in previous work
\cite{vonOppen98:4494,Mirlin98:1070}, and thus the inclusion of the effect of
``channeled orbits'', reproduces the typical
V-shaped resistance minimum at $\nu=1/2$ ($\bar{B}_{\mathrm{eff}}=0$) and yields
good overall agreement with experiment, consistent with the low-field 
magnetoresistance oscillations.
Including further the superposition of the original electric
superlattice and the induced effective magnetic superlattice, we can also
reproduce the asymmetry of the resistance curve around  $\bar{B}_{\mathrm{eff}}=0$. 

We are grateful to J. Smet for providing the unpublished data of
Fig.~\ref{smet_285} and for helpful discussions.
This work was supported by BMBF Grant No. 01BM622.

\bibliographystyle{prsty}

%\bibliography{/home/home2/zwersch/tex/literatur/zitate}

\begin{thebibliography}{10}

\bibitem{Jain89:199}
J.~K. Jain, Phys. Rev. Lett. {\bf 63},  199  (1989).

\bibitem{Halperin93:7312}
B.~I. Halperin, P.~A. Lee, and N. Read, Phys. Rev. B {\bf 47},  7312  (1993).

\bibitem{Fleischmann96:167}
R. Fleischmann, T. Geisel, C. Holzknecht, and R. Ketzmerick, Europhys. Lett.
  {\bf 36},  167  (1996).

\bibitem{Mirlin97:3717}
A.~D. Mirlin and P. W{\"o}lfle, Phys. Rev. Lett. {\bf 78},  3717  (1997).

\bibitem{vonOppen98:4494}
F. von Oppen, A. Stern, and B.~I. Halperin, Phys. Rev. Lett. {\bf 80},  4494
  (1998).

\bibitem{Smet98:4538}
J. Smet, K. von Klitzing, D. Weiss, and W. Wegscheider, Phys. Rev. Lett. {\bf
  80},  4538  (1998).

\bibitem{Beenakker89:2020}
C.~W.~J. Beenakker, Phys. Rev. Lett. {\bf 62},  2020  (1989).

\bibitem{Mirlin98:1070}
A.~D. Mirlin, P. W{\"o}lfle, Y. Levinson, and O. Entin-Wohlman, Phys. Rev.
  Lett. {\bf 81},  1070  (1998).

\bibitem{Menne98:1707}
R. Menne and R.~R. Gerhardts, Phys. Rev. B {\bf 57},  1707  (1998).

\bibitem{Simon98:91}
S.~H. Simon,  in {\em Composite Fermions}, edited by O. Heinonen (World
  Scientific, Singapore, 1999), p.\ 91.

\bibitem{Ye95:3013}
P.~D. Ye {\it et~al.}, Phys. Rev. Lett. {\bf 74},  3013  (1995).

\bibitem{Gerhardts96:11064}
R.~R. Gerhardts, Phys. Rev. B {\bf 53},  11064  (1996).

\bibitem{Mueller92:385}
J.~E. M{\"u}ller, Phys. Rev. Lett. {\bf 68},  385  (1992).

\bibitem{Beton90:9229}
P.~H. Beton {\it et~al.}, Phys. Rev. B {\bf 42},  9229  (1990).

\bibitem{Aronov94:16609}
A.~G. Aronov, A.~D. Mirlin, and P. W{\"o}lfle, Phys. Rev. B {\bf 49},  16609
  (1994).

\bibitem{SmetPrivat}
J.~H. Smet, private communications

\bibitem{footnote}
The insert of Fig.~\ref{mod_400} shows the measured low-field
resistance. We need extremely pronounced forward scattering
in order to reproduce the small measured resistance
and simultaneously the strong suppression of Weiss
oscillations at $B<0.05\,$T. The good agreement is of course limited by the
Shubnikov-de Haas oscillations which in these 
high-mobility samples (at $T=50\,$mK) set in at $B>0.2\,$T.

\end{thebibliography}
%\bibliographystyle{prsty}

\begin{figure}
\caption{Resistivity near filling factor $1/2$, $a=400\,$nm,
  $\bar{n}=1.82\cdot10^{11}\,\mathrm{cm}^{-2}$; thick dashed line: 
 experiment of Ref.\protect\cite{Smet98:4538}, 
 thin dash-dotted line: theory of Ref.~\protect\cite{Mirlin98:1070}
 (parameters: $\rho_0^{\mathrm{CF}}=270\,\Omega$ and
 $\epsilon_{\mathrm{CF}}=2.6$\%), 
 thin solid line: results of full calculation (including channeled
 orbits) with $p=6$ and $b=0$, thin dotted line: Beenakker 
 approximation for same parameters,
 thick solid line: full
 calculation for $\rho_0^{\mathrm{CF}}=650\,\Omega$, $\epsilon_{\mathrm{CF}}=3.5$\%,
  $p=2$ and $b=0.1$.  The inset shows the
 low-field magnetoresistance; dashed line: experiment, solid line: full
 calculation based on Ref.~\protect\cite{Menne98:1707} 
 for $\rho_0^{\mathrm{el}}=4.5\,\Omega$ and $\epsilon_{\mathrm{el}}=2.8$\%,  $b=0$ and
 $p=15$.  
 }\label{mod_400}
\end{figure}

\begin{figure}
\caption{Resistivity near filling factor $1/2$, $a=275\,$nm,
 $\bar{n}=1.98\cdot10^{11}\,\mathrm{cm}^{-2}$; thick dashed line: 
 experimental data, thick solid line: calculation for 
 $\rho_0^{\mathrm{CF}}=700\,\Omega$,
 and $\epsilon_{\mathrm{CF}}=3.8$\%, cross section parameters $p=2$ and $b=0.1$,
 as in Fig.~\protect\ref{mod_400}
 (thick lines), thin dotted line: Beenakker-type approximation (same
 parameters). The inset shows the
 low-field magnetoresistivity; dashed line: experiment, solid line: calculation
 for $\rho_0^{\mathrm{el}} =7.5\,\Omega$, $\epsilon_{\mathrm{el}}=3.6$\%,
  $b=0$, and $p=15$.
 }\label{smet_285}
\end{figure}

\onecolumn
\newpage
\Huge Fig.\ref{mod_400} \normalsize Zwerschke\\
\epsfig{file=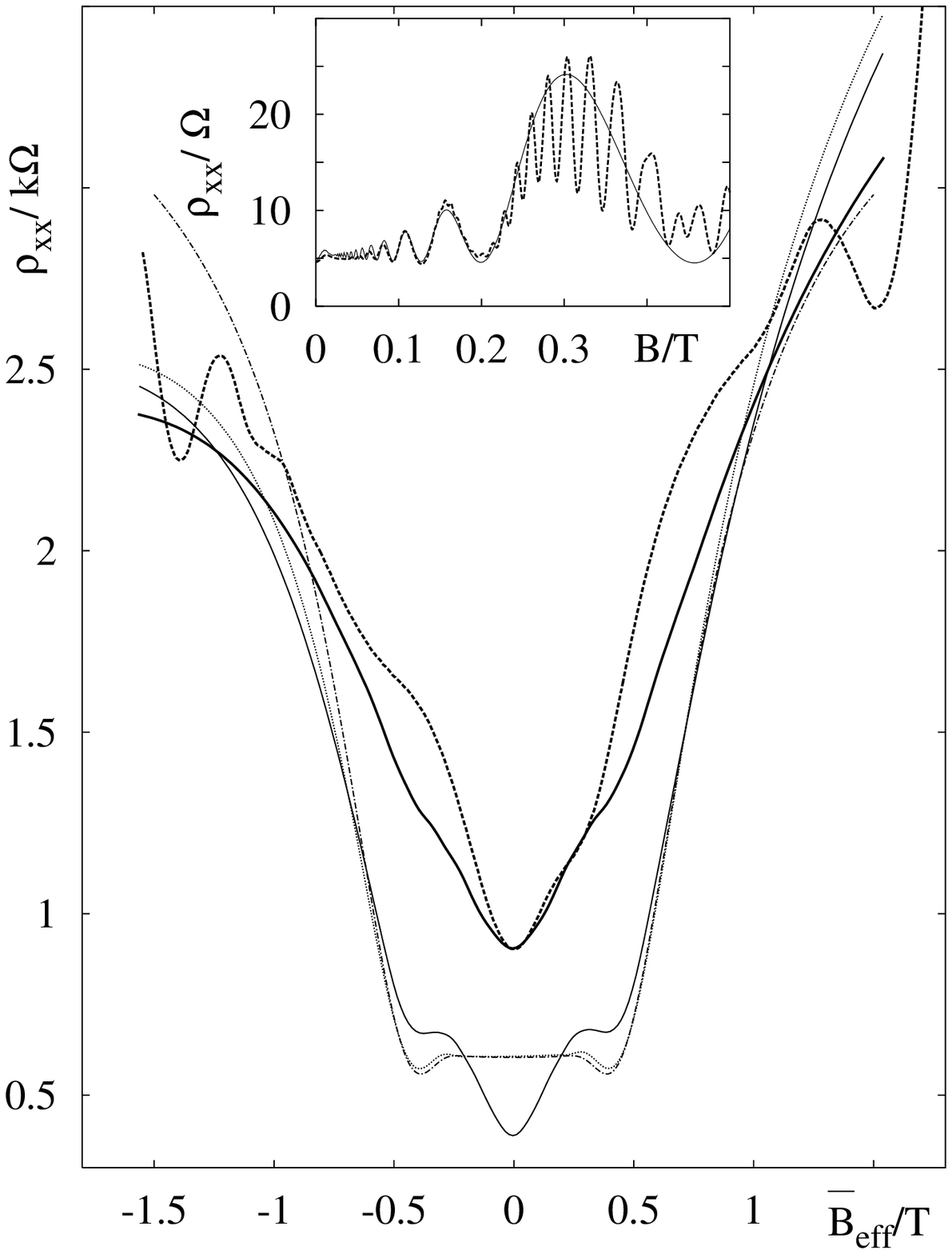}
\vfill
\newpage
\Huge Fig.\ref{smet_285} \normalsize Zwerschke\\
\epsfig{file=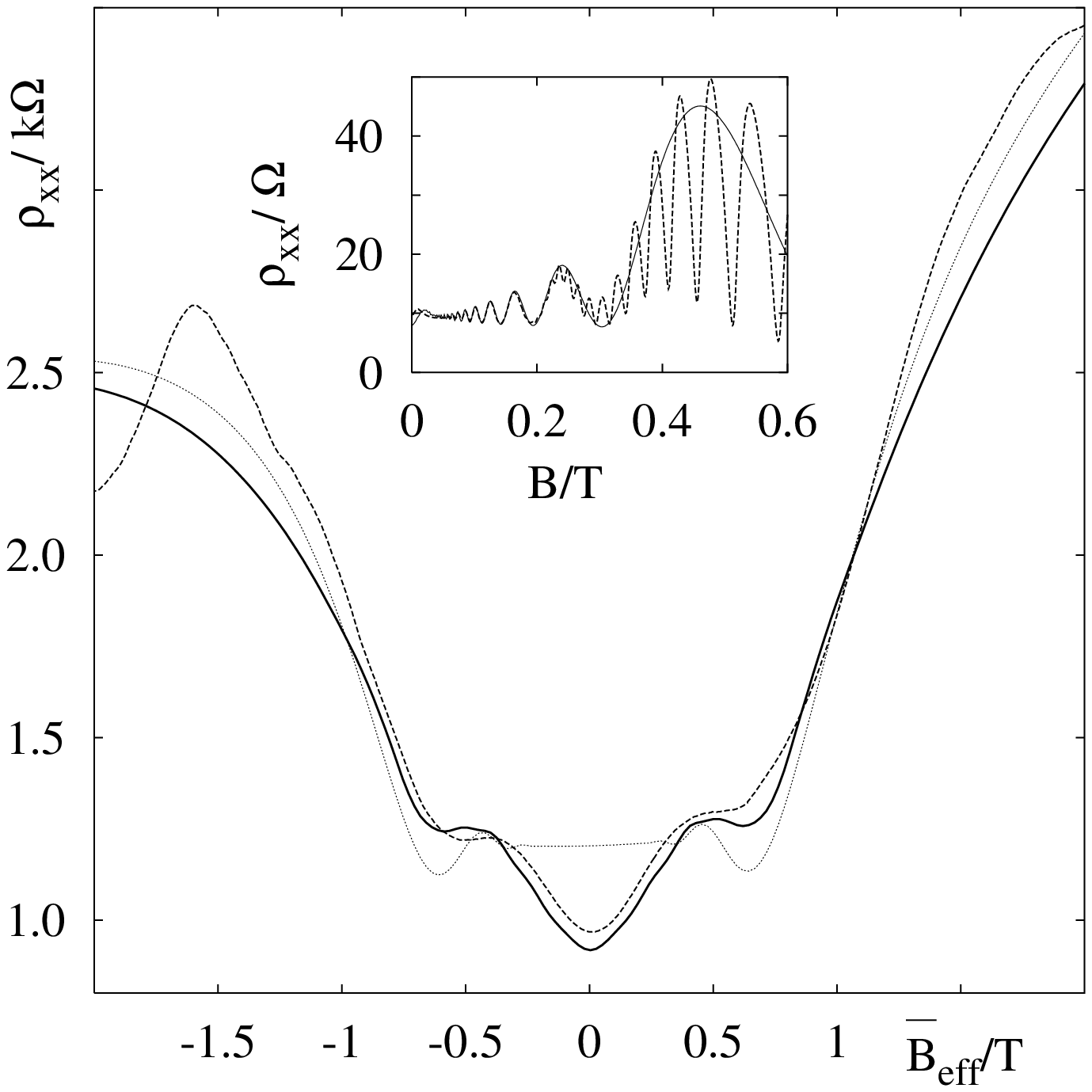}
\vfill
\end{document}